# Persistence of magnons in a site-diluted dimerized frustrated antiferromagnet.

M. B. Stone[1], A. Podlesnyak[1], G. Ehlers[1], A. Huq[1], E. C. Samulon[2], M. C. Shapiro[2], and I. R. Fisher[2]

*1 Neutron Scattering Science Division, Oak Ridge National Laboratory, Oak Ridge, Tennessee 37831, USA*

*2 Department of Applied Physics and Geballe Laboratory for Advanced Materials, Stanford University, California 94305, USA*

We present inelastic neutron scattering and thermodynamic measurements characterizing the magnetic excitations in a disordered non-magnetic substituted spin-liquid antiferromagnet. The parent compound $Ba_3Mn_2O_8$ is a dimerized, quasi-two-dimensional geometrically frustrated quantum disordered antiferromagnet. We substitute this compound with non-magnetic vanadium for the $S = 1$ manganese atoms, $Ba_3(Mn_{1-x}V_x)_2O_8$, and find that the singlet-triplet excitations which dominate the spectrum of the parent compound persist for the full range of substitution examined, x = 0.02 to 0.3. We also observe additional low-energy magnetic fluctuations which are enhanced at the greatest substitution values. These excitations may be a precursor to a low-temperature random singlet phase which may exist in $Ba_3(Mn_{1-x}V_x)_2O_8$.

## I. INTRODUCTION

Dimerized quantum disordered antiferromagnets (AFMs) serve as clean systems for experimentally testing theoretical models.[1,2,3] One approach to study these systems is to initially characterize the quantum ground state and its excitations, and then explore emergent phenomena by perturbing the system via external tunable parameters. One can then examine how these parameters change the intrinsic dynamics of the system. Beyond temperature, clean external parameters available to modify the ground state of such systems include applied magnetic fields,[4,5,6] and hydrostatic and uniaxial pressure.[7,8,9,10] One other approach to modifying the quantum states of these systems is to chemically perturb the Hamiltonian via magnetic or non-magnetic chemical substitution.[11,12,13,14]

Doping dependent experimental studies of AFMs have primarily used thermodynamic measurements to examine the creation of or the change in ordering temperature of a long-range magnetically ordered phase as a function of chemical substitution. In the case of the spin-Peierls (SP) compound $CuGeO_3$, the SP temperature decreases with initial substitution, and eventually a long-range ordered AFM phase develops with further substitution of magnetic sites.[15] The development of long range magnetic order and its temperature dependence for substitution on the magnetic site has also been examined in several systems with disordered groundstates and a finite energy gap or spin-gap to excitations. These include the quantum spin ladder $Sr(Cu_{1-x}Zn_x)_2O_3$,[16] the Haldane chain $Pb(Ni_{1-x}Mg_x)_2V_2O_8$,[17] and the coupled quantum dimer system $TlCu_{1-x}Mg_xCl_3$.[18] In these examples it is proposed that diluting the magnetic sites creates unpaired spins in the vicinity of the non-magnetic substituted site. These unpaired spins effectively extend the coupling of lower dimensional systems into three-dimensions eventually leading to long-range-order of the entire magnetic lattice.[19,20]

Ba$_3$Mn$_2$O$_8$ is an antiferromagnet composed of $S=1$ 3d$^2$ Mn$^{5+}$ spin-pairs or dimers arranged vertically on a hexagonal lattice.[33] Figure 1(a) and (b) illustrates the crystal structure and the intradimer and interdimer exchange interactions respectively.[33,24] Hexagonal layers of dimer sites are separated from one another along the crystallographic *c*-axis by 12-fold oxygen coordinated Ba sites. Initial thermodynamic measurements characterized the spin-gap and a linear sum of exchange constants.[21,22,36] The thermodynamic measurements indicated that the magnetic ground state of Ba$_3$Mn$_2$O$_8$ is a non-magnetic singlet phase; the antiferromagnetic dimer excitations are from the $S=0$ singlet ground state to the $S=1$ triplet and $S=2$ quintet states. Previous INS measurements of powders and single crystals directly measured the propagating triplet modes of the coupled dimers and determined the significant exchange interactions and their magnitude.[23,24,25] These measurements found that the intra- and inter-layer exchange enhances the two-dimensional (2D) nature of the magnetic lattice. There are also two low-temperature high field regions of long range magnetic order in Ba$_3$Mn$_2$O$_8$. These are essentially field-induced condensates of the triplet and quintet states of the dimer network. The value of the spin-gaps, singlet-triplet $\Delta \approx 1$ meV and singlet-quintet $\Delta \approx 3.4$ meV, allows one to experimentally access both critical fields of these phases.[36,21,26,22] The influence of the zero-temperature quantum critical point (QCP) on thermodynamic parameters has also been measured in detail.[22]

Ba$_3$(Mn$_{1-x}$V$_x$)$_2$O$_8$ forms for all values of 0≤x≤1. In this compound, $S=0$ 3d$^0$ V$^{5+}$ ions are substituted for $S=1$, 3d$^2$ Mn$^{5+}$ ions. Because the vanadium and manganese ions have the same valence no charge is added to the system. However, substitution of non-magnetic atoms affects the magnetic ground state, by breaking singlet dimers and affecting both the excited triplet band while also creating $S=1$ Mn moments which are not as strongly exchange coupled as those of the $x=0$ dimers.

In this article, we present an experimental study of non-magnetic substitution of a model quantum system $Ba_3(Mn_{1-x}V_x)_2O_8$. In addition to magnetic susceptibility, we use inelastic neutron scattering (INS) to probe the magnetic excitations directly. We find that the dimer excitations of the parent compound[23,24] remain remarkably stable even for the greatest levels of substitution that we have examined. Such excitations have also been directly probed in other chemically substituted systems using INS, however, a long range ordered phase is very often also observed.[27,12,28] Our measurements indicate no sign of a transition to short or long range magnetic order down to $T = 1.8$ K. One of the most dramatic changes in the spectrum is the appearance of low-energy fluctuations for the greatest levels of substitution. As the dimer singlet-triplet (*i.e.* triplet) excitations are depopulated with non-magnetic substitution, these additional low energy fluctuations increase in number. These excitations may be the precursor of a spin glass state,[29] or possibly the high temperature manifestation of a random singlet phase, as has been suggested for this system by Samulon *et al*.[30]

## II. EXPERIMENTAL METHODS

Powder samples of $Ba_3(Mn_{1-x}V_x)_2O_8$ were synthesized via a solid state reaction following the procedure previously described for $Ba_3Mn_2O_8$.[23,36] Powder samples were produced for x = {0,0.02,0.035,0.05,0.2,0.3}. Stoichiometric amounts of $BaCO_3$, $Mn_2O_3$ and $V_2O_5$ were calcined under flowing oxygen at 900 °C for 30 hours, and the resulting powder was then reground and sintered in temperatures up to 1050 °C for 70 hours. $x = 0.02$ and $x = 0.3$ samples were examined for phase purity using a laboratory Cu anode X-ray powder diffractometer. No additional phases or unreacted materials were observed for V concentrations of x=0.02 and x = 0.3.

Low field magnetic DC susceptibility measurements were performed on powder samples using a commercial SQUID magnetometer. Measurements were performed between $T = 1.8$ K and $T = 300$ K in an applied magnetic field of $\mu_0 H = 0.1$ T. The normalized magnetic susceptibility was determined by dividing the magnetization, $M$, by the applied magnetic field, $\chi = M/\mu_0 H$. No phase transitions associated with potential metal oxide impurity phases were observed in any of the samples.

High resolution neutron powder diffraction measurements were performed for a range of dopings. These were performed using the Powgen time-of-flight diffractometer at the Spallation Neutron Source (SNS). Diffraction patterns were acquired on powder samples at room temperature. Samples consisted of approximately two grams of powder contained in vanadium sample cans. Samples were measured for approximately 2.5 hours using 1.066 Å wavelength neutrons.

Inelastic neutron scattering spectra were measured using the CNCS instrument at the SNS. Spectra were acquired using powder samples in aluminum sample cans under He exchange gas. The sample environment consisted of a He-flow cryostat. A cadmium mask was applied directly to the aluminum sample can to reduce multiple scattering and aluminum scattering from the cryostat. The powder samples were measured with $E_i = 3.2$ meV incident energy neutrons using the standard resolution configuration of the instrument. This yielded a full width at half maximum (FWHM) elastic energy resolution of 0.0736(5) meV[31] as determined from the incoherent elastic scattering of a vanadium standard. Sample masses for different substitution values were each approximately 5 grams. The data were individually normalized to the calculated moles of $Mn^{5+}$ spin in each sample.

III. RESULTS

We have characterized the crystal structure of four different substitution values for Ba$_3$(Mn$_{1-x}$V$_x$)$_2$O$_8$, x = {0.035, 0.1, 0.2, 0.3}, via powder neutron diffraction. We observe no sign of a structural phase transition as a function of vanadium doping. Figure 2 shows the measured scattering intensity for the x = 0.2 sample as a function of d-spacing. Figure 1(c) shows a small portion of these diffraction data as a function of d-spacing. There is a small but clear shift in lattice constants as a function of doping. We refined the diffraction data both fixing the atomic positions of all but the Mn and V sites based upon the x=0 structure, and allowing these positions to vary.[32] The quality of refinement did not improve when the atomic positions were allowed to vary. An example of the refined diffraction measurement is shown as a solid line in Fig. 2, and agrees very well with the measurements. The refined lattice constants are shown in Fig. 1(d) and (e) as a function of vanadium substitution. For comparison, we show the literature x = 0 values from Ref. 33 as open symbols. For small x, the results agree very well with the literature values. As the vanadium doping increases, the *a* lattice constant slowly lengthens, the *c* axis gets shorter, and the ratio *c/a* becomes smaller. Laboratory X-ray measurements were performed for x = 0.02 and 0.3. These data are also shown in Fig. 1(d) and (e). No change in lattice constant was observable from the X-ray measurements for the x = 0.02 doping level. The trend as a function of vanadium substitution agrees well with the room temperature lattice constants determined for Ba$_3$V$_2$O$_8$: *a* = 5.7733(14), c = 21.339(10) and a = 5.7714(9), c=21.2480(30) from Refs. 34 and 35 respectively. The overall changes in lattice constant are small as a function of substitution: less than half a percent for the *a*-axis and less than 0.1 percent for the *c*-axis. This implies that the exchange constants between Mn$^{5+}$ sites are of a very similar magnitude for all samples in this range of vanadium substitution. Consequently, any effects observed in the magnetic properties will be primarily due to intrinsic effects of non-magnetic substitution and not due to changes in exchange constants due to changes in the crystal structure.

Figure 3 displays powder magnetic susceptibility as a function of temperature for $Ba_3(Mn_{1-x}V_x)_2O_8$ with x = {0,0.02,0.035,0.05,0.2,0.3}. The curves for small values of x have a characteristic gradual decrease as temperature is decreased below $T$ = 20 K. This behavior is due to the quasi-2D spin-gap phase becoming well defined as the temperature decreases. For the x = 0 and x ≤ 0.035 samples, there is a roll-over at approximately 20 K. As substitution increases, this roll-over is gradually dominated by a rapidly increasing component of the susceptibility at low-temperatures. This low temperature increase is induced by unpaired Mn spins which are created by vanadium sites breaking the spin dimers which define the spin gap phase at lower temperatures. Figure 3(b) shows the inverse susceptibility as a function of temperature for the different substitution values. There are separate power-law dependences at both high and low temperatures.

Figure 4 illustrates the measured INS spectra for x = 0.02 and x = 0.3 of $Ba_3(Mn_{1-x}V_x)_2O_8$. The x = 0.02 data appear nearly identical to the x = 0 powder measurements performed previously.[23] There is a single gapped excitation between approximately $\hbar\omega$ = 1 meV and 2.5 meV energy transfer. This is the magnon excitation associated with the transition from the singlet ground state to the triplet of states of the $S$ = 1 spin dimer. The intensity is strongly modulated as a function of wave vector transfer, $Q$, with a peak at $Q$ = 1 Å$^{-1}$. The quasi-2D nature of the exchange coupling in the compound results in a characteristic smeared "V" in the powder INS spectrum as a function of energy and wave vector transfer. This is still evident in the x = 0.02 data in Fig. 4(a). The x = 0.3 data still have this characteristic shape albeit less well defined. The wave vector dependent intensity modulation is nearly identical to that observed for the x = 0.02 data. This implies that the strongest exchange interactions are between the same $Mn^{5+}$ moments as in the x = 0.02 samples and likewise the x = 0 material, i.e. the strong dimer bond is unaffected by non-magnetic substitution. The gap in the spectrum is slightly larger for the x = 0.3 sample. The greatest difference in these spectra is the existence of additional low-energy fluctuations in the x = 0.3 sample. These quasielastic excitations are localized around elastic energy transfers and generally decrease in intensity as wave vector increases, implying a magnetic origin. In addition from examining the elastic scattering channel, we observe no structural phase transition or transition to AFM long-range order as a function of temperature and vanadium substitution.

## IV. DISCUSSION

We compare the magnetic susceptibility as a function of temperature, *T*, and doping to a simple sum of the spin-dimer contributions and the free-spin contribution:

$$\chi = \left(\frac{C}{T-\Theta}\right) + \frac{\chi_d}{1+\lambda\chi_d} + \chi_0, \tag{1}$$

where

$$\chi_d = \frac{2N\beta g^2 \mu_B^2 (1+5e^{-2\beta J})}{3+e^{\beta J}+5e^{-2\beta J}}, \tag{2}$$

$$\lambda = \frac{3[J_1+J_4+2(J_2+J_3)]}{Ng^2\mu_B^2} = \frac{3J'}{Ng^2\mu_B^2}, \tag{3}$$

$\beta = \frac{1}{k_B T}$, *g* is the Lande-*g* factor, $\mu_B$ is the Bohr magneton, and *N* is the number of dimers per mole. In Eq. (1), $\left(\frac{C}{T-\Theta}\right)$ is a Curie-Weiss law contribution from weakly coupled moments. This portion of the magnetic susceptibility is included to describe the low-temperature behavior of the more highly vanadium substituted samples. The second contribution in Eq. (1) is a mean field expression for the dimer contribution to the magnetic susceptibility, including the intradimer exchange constant *J* and inter-dimer, exchange constants $J_1$, $J_2$, $J_3$ and $J_4$.[36] Equation (3) represents the linear combination of the inter-dimer exchange constants with the effective exchange $J'$. An initial fit is performed to the data for the parent compound, x = 0. We then fix the values of the exchange constants and Lande-$g$ factor determined from the x = 0 sample in the magnetic susceptibility fits for the x ≠ 0 samples. This is reasonable, given the very small change in lattice constants for this range of substitution, and the lack of a significant change to the magnon spectrum defined by the exchange constants $J$ and $J'$. The lines shown in Fig. 3 correspond to these fits. These results agree very well with the measurements over a large range of temperature and substitution level. Figure 5(a) shows the variation in the unpaired spin concentration, $C$ from Eq. (1), as a function of x. The x = 0.035 data have a larger Curie-Weiss response at low-temperature than expected based upon the measurements of nearby substitution values in the series of samples examined. Single crystal samples grown from the x = 0.035 powder starting material were measured to have appropriate doping, indicating that there may be some unsintered material in the x = 0.035 concentration sample contributing to the low-temperature magnetic susceptibility. There is a linear increase (fitted line shown in Fig. 5(a)) in the free-spin contribution to the magnetic susceptibility as a function of x, revealing the additional paramagnetic response in substituted samples. We will show that this linear increase agrees well with the increase in the additional magnetic fluctuations observed in the INS spectra at small energy transfers as shown in Fig. 4(b).

We first characterize the INS measurements of the gapped portion of the magnetic spectrum and then we examine the nature of the low-energy fluctuations. Our measurements of the x = 0 compound found that the wave vector dependent scattering intensity of the triplet excitation agreed very well with the scattering function $\mathcal{S}(\mathbf{Q},\hbar\omega)$ for isolated spin-1 dimers:

$$\mathcal{S}(\mathbf{Q},\hbar\omega) = \frac{4e^{2J\beta}[1-\cos(\mathbf{Q}\cdot\mathbf{d})]}{e^{2J\beta}+3e^{J\beta}+5e^{-J\beta}}\delta(\hbar\omega - J). \tag{4}$$

In Eq (4), *J* is the exchange constant of the dimer and **d** is the vector between the coupled spins of the dimer. The powder averaged first frequency moment of the *T* = 0 scattering function yields information regarding the length of the dimer bond *d*:

$$\hbar\langle\omega\rangle \equiv \int_{-\infty}^{\infty} \int \frac{d\Omega d\omega}{4\pi} \hbar\omega(\mathbf{Q}) \, S(\mathbf{Q}, \hbar\omega) \propto |F(Q)|^2 \left[1 - \frac{\sin(Qd)}{Qd}\right], \quad (5)$$

where *F(Q)* is the magnetic form factor of the magnetic ion. The incident energy used for our measurement was chosen to only measure energies to the top of the triplet excitation while preserving energy resolution in order to observe low-energy excitations in the vicinity of the incoherent elastic scattering. The kinematic constraints of this choice limits the range of wave vector transfers, such that only the first maximum of Eq (5) is clearly observed in our measurements of $Ba_3(Mn_{1-x}V_x)_2O_8$. In Fig. 6, we plot the integrated scattering intensity for energy transfers between 0.5 and 2.9 meV for the *T* = 1.8 K measurements. All of these data have a peak at approximately the same wave vector transfer indicating that the dominant dimer bond is independent of non-magnetic substitution. We numerically calculate the peak in the first moment spectrum as a function of the dimer length, *d*, using Eq (5). This relationship is shown as the inset of Fig. 6. Integrating over the entire triplet mode results in a value which is proportional to the first moment described in Eq. (5). Because of the limited *Q* range of the measurement, we perform a simultaneous Gaussian fit to the data in Fig. 6 using a common background and width for all lineshapes. The corresponding value of *d* based upon the peak location is plotted in Fig. 5(b) as a function of x. The determined values are independent of vanadium substitution and slightly less than that of the parent compound: for vanadium substituted samples the average value of *d*=3.964 Å while for the previously measured x=0 sample *d* = 4.073(7) Å.[23] The room temperature value of *d* based upon the parent compound crystal structure is *d* = 3.985 Å. The dashed line in Fig. 5(b) is a calculation of the crystallographic $d$ value based upon a smooth interpolation of the $c$-axis lattice constant data shown in Fig. 1(d). It is clear that the magnetic dimer contributing to the singlet-triplet excitation in $Ba_3(Mn_{1-x}V_x)_2O_8$ does not change as a function of x.

Figure 7 shows the INS momentum-integrated intensity as a function of energy transfer for the $T = 1.8$ K measurements for different substitution values x. These data are integrated between wave vectors $Q = 0.6$ and 1.7 Å$^{-1}$. The triplet excitation is the broad peak between one and approximately 2.5 meV energy transfer. One can also see additional paramagnetic excitations developing at low-energy transfers as the vanadium substitution is increased. x=0 single crystal measurements have very accurately determined the gap, $\Delta = 1.081$ meV, and upper maximum of the dispersion, 2.78 meV.[24] We use these values to parameterize the zero-field data shown in Fig. 7 by finding the ratio of inelastic intensity to the peak intensity at the spin-gap and maximum dispersion energy transfers. The gap and bandwidth of the vanadium substituted spectra are determined based upon the energy transfers for the intensity ratios of the x = 0 data. The spin-gap $\Delta$ (simple linear fit shown in Fig. 5(c)) increases with doping, and the bandwidth decreases slowly with doping (shown in Fig. 5(d)). As the coupled dimers are substituted non-magnetically, the triplet excitations are no longer able to propagate as freely and the amount of dispersion decreases i.e. approaches the dispersionless isolated dimer limit.

To examine the low-temperature quasi-elastic peaks, we parameterize the energy dependent scattering intensity in Fig. 7 as a sum of Gaussian and Lorentzian lineshapes. The inelastic triplet excitation is parameterized as a sum of two Gaussians, and the elastic incoherent scattering is parameterized as a single Gaussian peak. The low-energy quasi-elastic scattering is parameterized as a Bose-factor normalized anti-symmetric Lorentzian function (see Eq. 12 of Ref. 23). This lineshape was then convolved with the energy dependent resolution function of the spectrometer. The ratio of the integrated intensity of the quasi-elastic scattering to the inelastic scattering is shown in Fig. 5(e) as a function of x. As substitution increases, the relative amount of quasi-elastic scattering increases. These paramagnetic fluctuations are most likely due to the increasing number of clusters of broken and weakly coupled dimer pairs in the lattice, and reflect the fact that there are several additional relevant interdimer exchange constants in this system. In Fig. 5(e) the dashed line is the calculated fraction of broken dimer bonds as a function of x. The fraction of quasi-elastic scattering exceeds the number of half magnetically occupied dimers for larger values of vanadium substitution. This indicates that the quasi-elastic scattering is not solely due to broken dimer pairs, but there is likely a contribution from other spin-spin interactions in the lattice. For the range of x examined, the scattering intensity ratio is nearly linear; a linear fit is shown as a solid line in Fig. 5(e). We also note that a similar linear increase in low-temperature paramagnetic fluctuations was observed in the powder magnetic susceptibility data as illustrated in the Curie-Weiss constant $C$ in Fig. 5(a).

We also examine the temperature dependent scattering for both the triplet excitation and the quasi-elastic scattering. Figure 8(a) shows the integrated triplet scattering intensity as a function of temperature and vanadium substitution integrated between wave vector transfer $0.5 < Q < 1.5$ Å$^{-1}$ and energy transfer $0.5 < \hbar\omega < 2.9$ meV. We simultaneously fit these data to the temperature dependence described in Eq. (4) for the triplet excitation with a common value of dimer exchange, $J$, and a common constant background term. The solid lines in Fig. 8(a) are the results of this comparison and indicate that this provides a good parameterization of the measured results. The resulting $J=2.6(2)$ meV is larger than the x = 0 value of $J_0$, but a reasonable approximation of the midpoint of the triplet spectrum. Even at elevated temperatures, the triplet mode of the non-magnetic substituted compound is well approximated by the simple dimer model.

Figure 8(b) shows the inverse dynamic susceptibility as a function of temperature for the different values of x examined. The data were integrated between wave vector transfer $0.5 < Q < 1.5$ Å$^{-1}$ and energy transfer $0.25 < \hbar\omega < 0.5$ meV. The integrated intensity was then normalized by the Bose occupation factor, $[1 - \exp\left(-\frac{\hbar\omega}{k_B T}\right)]^{-1}$, using the mean energy transfer of the integrated data to yield the dynamic susceptibility, $\chi_{INS}(T)$. The inverse dynamic susceptibility increases linearly as a function of temperature for all of the vanadium substitutiosn of Ba$_3$Mn$_{2(1-x)}$V$_x$O$_8$ including the x = 0 compound, (linear fits are shown in Fig 8(b)). This behavior is essentially the Curie-Weiss magnetic susceptibility response of the low-energy magnetic fluctuations. Although the lack of an absolute normalization prevents a comparison of the x-axis intercepts, i.e. the Weiss temperature, we can compare the slope of the fitted lines. For the inverse susceptibility, the inverse slope is a linear function of the density of moments contributing to the dynamic susceptibility. The slope of the lines decreases with increasing x indicating that as the non-magnetic vanadium substitution increases, additional magnetic excitations are being populated at lower energy transfers. This description agrees very well with the measured DC magnetic susceptibility shown in Fig. 3 and the fitted parameters determined from the low-temperature Curie-Weiss contribution. In Fig. 9, we plot the determined inverse slope of the data in Fig. 8(b) as a function of the determined value of *C* from the low-temperature Curie-Weiss law fit of the DC magnetic susceptibility in Fig. 3. The results agree well with one another; a linear fit forced to pass through the origin is shown in Fig. 9. The low energy fluctuations are indeed magnetic in origin.

The quasi-two-dimensional Shastry-Sutherland[1] frustrated quantum AFM $SrCu_2(BO_3)_2$ [37,38] has also been examined when non-magnetic Mg sites are substituted for the $S = 1/2$ Cu sites, $SrCu_{2-x}Mg_x(BO_3)_2$.[39,40] For this case, single crystal INS measurements found that upon non-magnetic substitution additional lower-energy excitations develop below the spin-gap energy concomitantly with a broadening of the triplet mode.[39] These effects are described as being due to unpaired dimer spins being coupled to neighboring triplet states.[41,42] The quasi-elastic excitations we observe developing in $Ba_3(Mn_{1-x}V_x)_2O_8$ are likely analogous to the additional spin excitations observed in $SrCu_{2-x}Mg_x(BO_3)_2$; however it is very likely that the reduced energy scale for the additional spin excitations in $Ba_3(Mn_{1-x}V_x)_2O_8$ is due to the additional interdimer coupling both within the *ab* plane and along the *c*-axis of the $Ba_3(Mn_{1-x}V_x)_2O_8$ structure. We note that the dispersion in $SrCu_2(BO_3)_2$ is very week within the plane and perpendicular to the plane due to the higher degree of magnetic frustration and isolated planar structure in this compound. Although we observe a small reduction in dispersion upon non-magnetic substitution, the large dispersion of $Ba_3Mn_2O_8$ and the effects of powder averaging do not allow us to place limits on the triplet lifetimes as was done for $SrCu_2(BO_3)_2$. Single crystal INS measurements of $Ba_3(Mn_{1-x}V_x)_2O_8$ would be very helpful to determine if there are any modes within the x = 0 triplet band similar to those observed in $SrCu_{2-x}Mg_x(BO_3)_2$.

V. CONCLUSIONS

Recently, there have been additional experimental results examining both magnetic and non-magnetic substitution of the $Mn^{5+}$ sites for $Ba_3Mn_2O_8$.[43] The non-magnetic substituted samples of Manna *et al.* were also based upon vanadium substitution of $Ba_3(Mn_{1-x}V_x)_2O_8$ with x = {0.25, 0.5, 1}. Their conclusions for the x = 0.25 and 0.5 samples were that the doping diminishes the formation of the antiferromagnetic dimer and thus eliminates the singlet-triplet excitations. Our INS measurements clearly show otherwise. Rather, the additional quasi-elastic magnetic fluctuations which develop with vanadium doping mask the signature of the singlet-triplet excitations in the magnetic susceptibility measurements for larger values of x.

In contrast to many other non-magnetically substituted gapped systems, no evidence for long range order was found in $Ba_3(Mn_{1-x}V_x)_2O_8$. Rather, vanadium substitution creates additional quasi-elastic excitations. These excitations may also be a high temperature manifestation of a random singlet phase,[44,45,46,47] as was proposed by Samulon *et al.* following their low temperature thermodynamic studies of this system.[30] The random singlet phase is marked by formation of local singlets without long range order or broken time reversal symmetry in contrast to the archetypal disordered ground state without long range magnetic order, the spin glass. A spin glass suffers a cooperative phase transition to a metastable state with broken time reversal symmetry, whereas a random singlet phase exhibits neither a phase transition nor broken time reversal symmetry. Certainly the fact that $Ba_3Mn_2O_8$ has a distribution of exchange energies available to each dimer spin makes this a reasonable possibility. Likewise, the temperature range over which Samulon *et al.* observes behavior indicative of a random singlet phase agrees well with the temperature and energy range of our observed quasi-elastic excitations. Further high resolution INS measurements using single crystals of intermediate values of x for $Ba_3(Mn_{1-x}V_x)_2O_8$ have the potential to determine the nature of the groundstate corresponding to the quasi-elastic excitations and the triplet excitations. Such measurements would also potentially be able to probe the dispersion of the quasi-elastic excitations at low-temperatures. Examining the temperature dependence of the low-energy spin-correlations may also provide further characterization of a potential random-singlet phase.[48]

ACKNOWLEDGEMENTS

M. B. S. thanks M. Lumsden, I. Zaliznyak and C. Batista for useful discussions. A portion of this research at Oak Ridge National Laboratory's Spallation Neutron Sourcewas sponsored by the Scientific User Facilities Division, Office of Basic Energy Sciences, U. S. Department of Energy. Work at Stanford was supported by the Department of Energy, Office of Basic Energy Sciences, under contract DE-AC02-76SF00515.

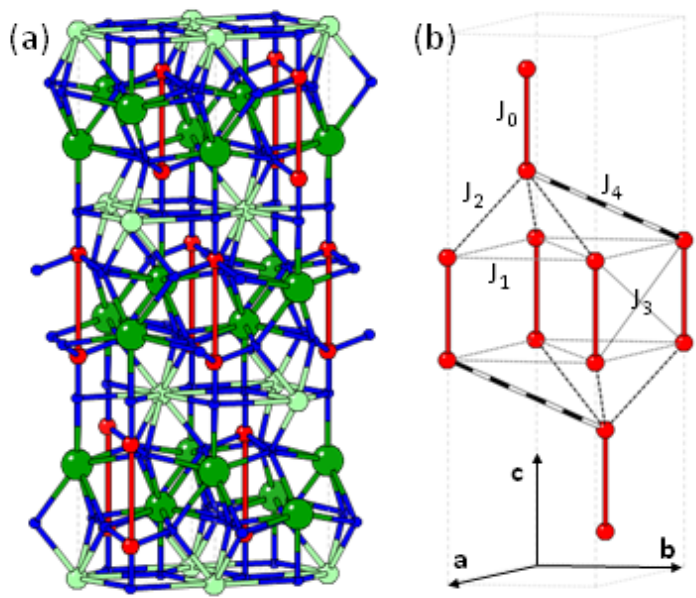

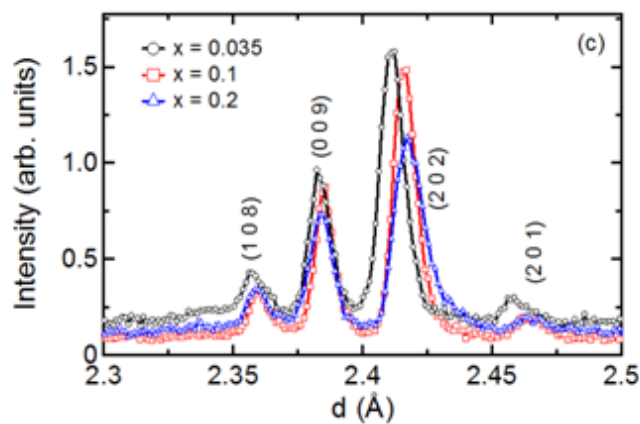

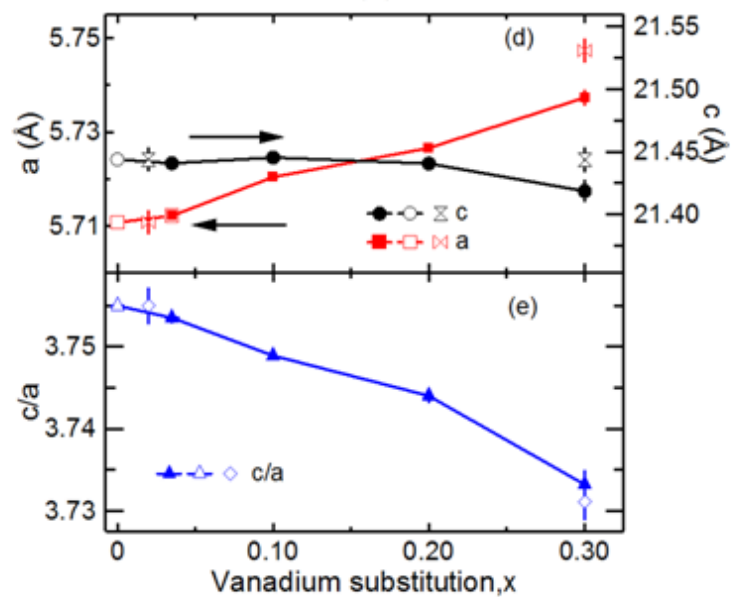

Figure 1. (color online) (a) Crystal structure of $Ba_3Mn_2O_8$.[33] Red circles represent Mn atoms, light green and dark green circles represent Ba atoms (12-fold and 10-fold coordinated respectively), and blue circles represent oxygen atoms. (b) $Mn^{5+}$ magnetic ions and the significant exchange interactions found for $Ba_3(Mn_{1-x}V_x)_2O_8$.[24] (c) Neutron powder diffraction intensity as a function of d-spacing for three values of x. Indicees of Bragg peaks are indicated for the peaks shown. (d) *a* and *c* lattice constants as a function of vanadium substitution in $Ba_3(Mn_{1-x}V_x)2O_8$. Open symbols for x = 0 are the literature (Ref. 33) values. Open symbols for x = 0.02 and 0.3 are from laboratory X-ray measurements. Vertical axis ranges were chosen to display lattice constants over the same range of $\Delta a/a$ as $\Delta c/c$. (e) *c/a* ratio as a function of vanadium substitution. Lattice constants were refined from time-of-flight neutron powder diffraction measurements (solid symbols) as described in the text. Error bars of neutron scattering data are shown as the standard deviation multiplied by five times the reduced chi squared value of the refinement. Measurements were performed at room temperature.

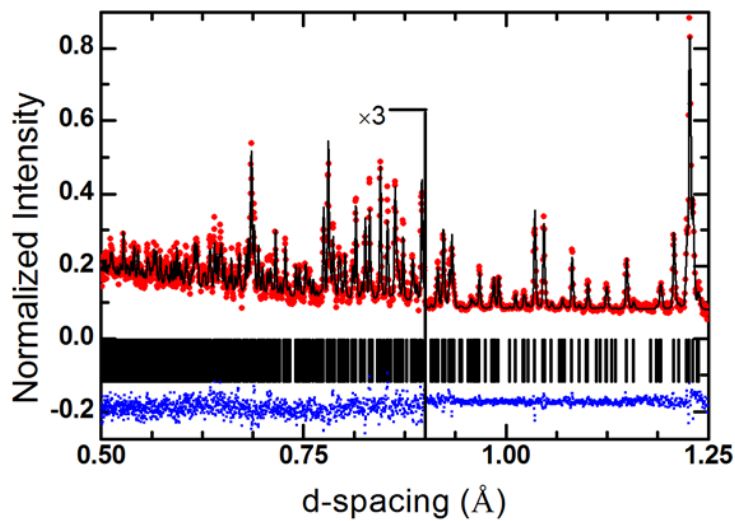

Figure 2. Rietveld refinement of $Ba_3(Mn_{1-x}V_x)2O_8$ for x = 0.2. Red circles are the measured scattering intensity. Solid black line is the refined diffraction pattern as described in the text. Solid vertical lines represent the location of individually index Bragg reflections. Blue squares at the bottom of the figure represent the difference in measured and refined scattering intensity. Data were measured at room temperature using the Powgen instrument. Refinement details are described in the text.

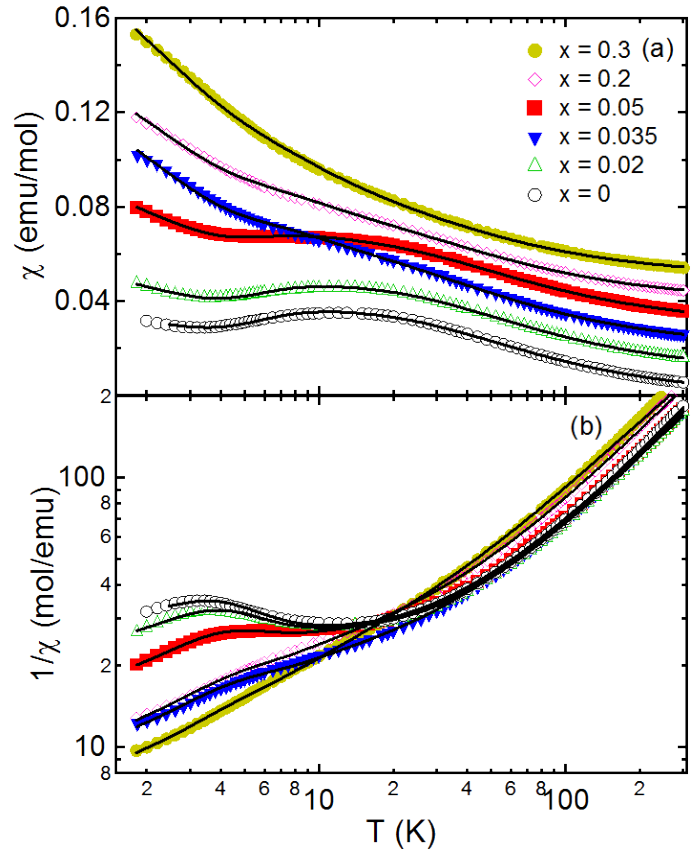

Figure 3. (color online) Magnetic susceptibility as a function of temperature for different concentrations of vanadium in $Ba_3(Mn_{1-x}V_x)_2O_8$. Data are normalized per mole of Mn spin. Solid lines are individual fits to a correlated spin (Curie-Weiss) plus $S = 1$ dimer model as described in the text. Data in (a) have been offset vertically in multiples of 0.01 emu/mol for presentation. Data in (b) are plotted as the inverse magnetic susceptibility on a log-log scale to emphasize power-law behavior at high and low temperatures.

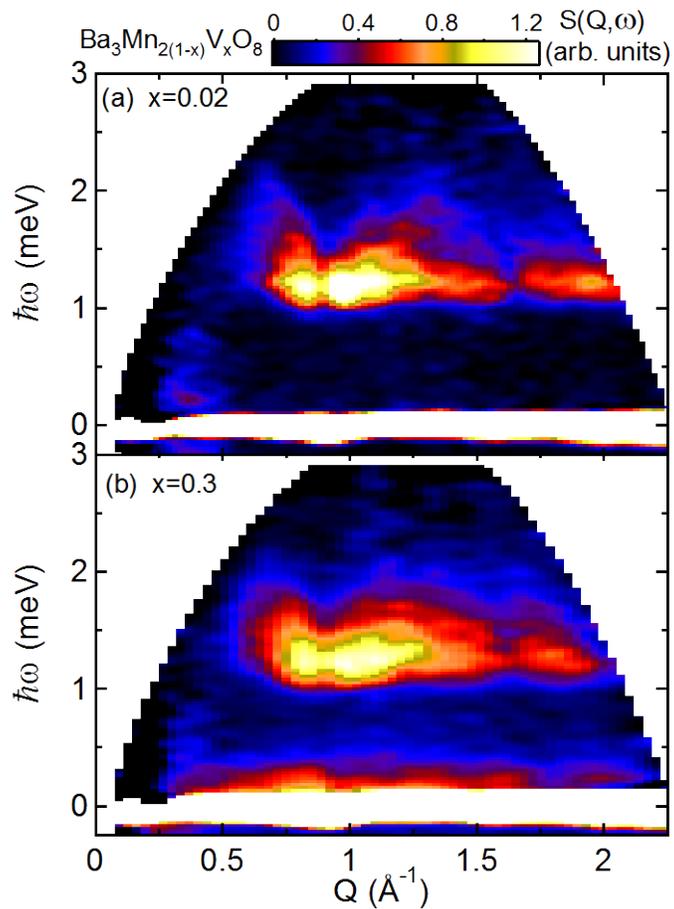

Figure 4. (color online) Inelastic neutron scattering intensity as a function of energy and wave vector transfer for $Ba_3(Mn_{1-x}V_x)_2O_8$ with (a) x=0.02 and (b) x=0.3. Data were acquired at T=1.8 K with 3.1 meV incident energy neutrons.

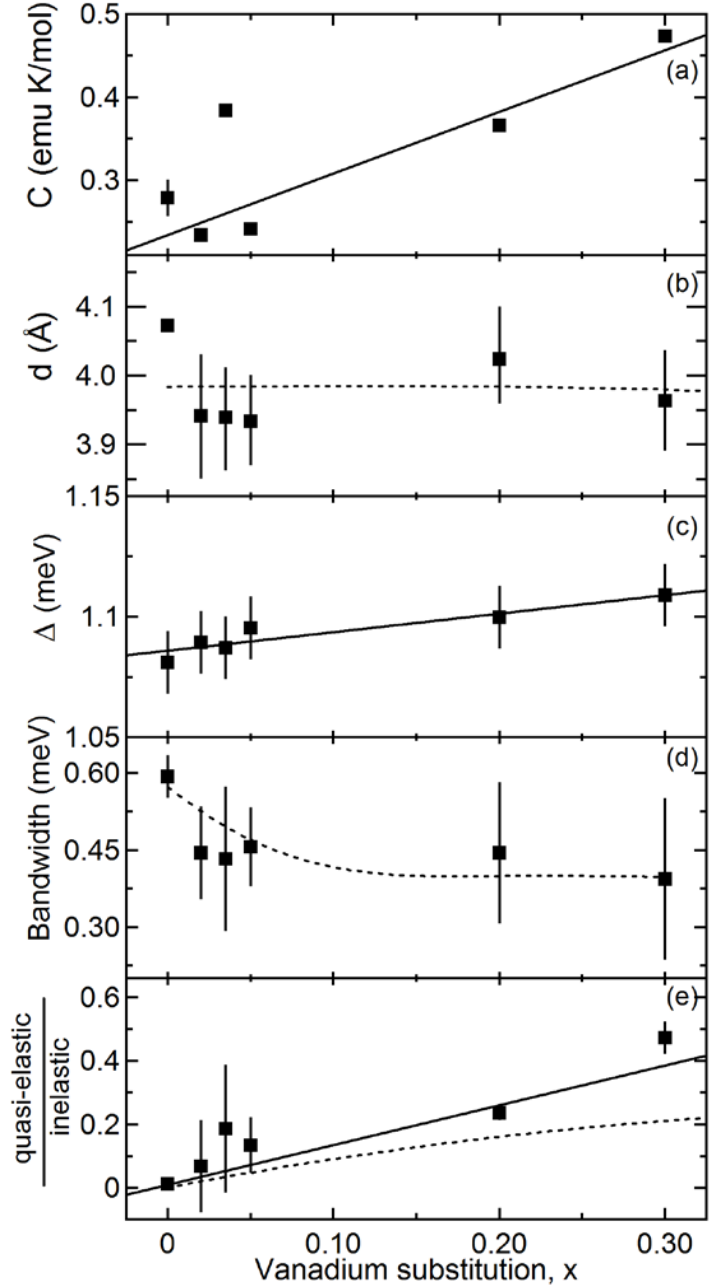

Figure 5. Parameters as a function of vanadium substitution x, of $Ba_3(Mn_{1-x}V_x)_2O_8$: (a) $C$ value determined from low-temperature powder magnetic susceptibility, (b) dimer length $d$, (c) Spin-gap, (d) bandwidth and (e) the ratio of the integrated quasi-elastic scattering intensity to the integrated inelastic scattering intensity. All parameters determined at 1.8 K. Solid lines are fits as described in the text. Dashed line in (b) is the calculated value of $d$ based upon the vanadium substitution dependent lattice constant shown in Fig. 1(a). Dashed line in (d) is a guide to the eye. Dashed line in (e) is the theoretical fraction of unpaired spins as a function of vanadium substitution.

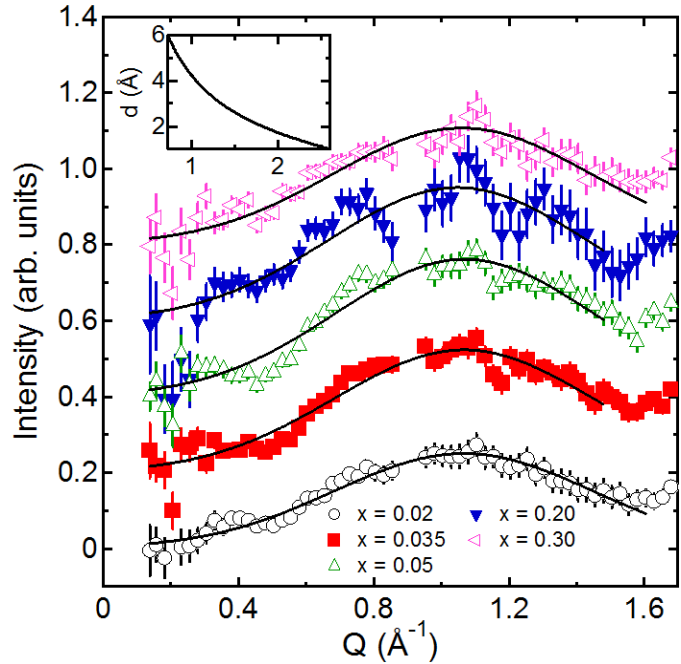

Figure 6. (color online) Wave vector dependence of scattering intensity for different vanadium substitutions, x, in $Ba_3(Mn_{1-x}V_x)_2O_8$ for $T = 1.8$ K measurements. Data correspond to integrating over the dimer excitation in the energy range of $0.5 < \hbar\omega < 2.9$ meV. Data have been offset vertically in multiples of 0.2 units. Solid lines are simple Gaussian fits with a sloping background. Inset shows the distance between dimer pairs plotted against the fitted peak in wave vector transfer for the dimer form-factor discussed in the text.

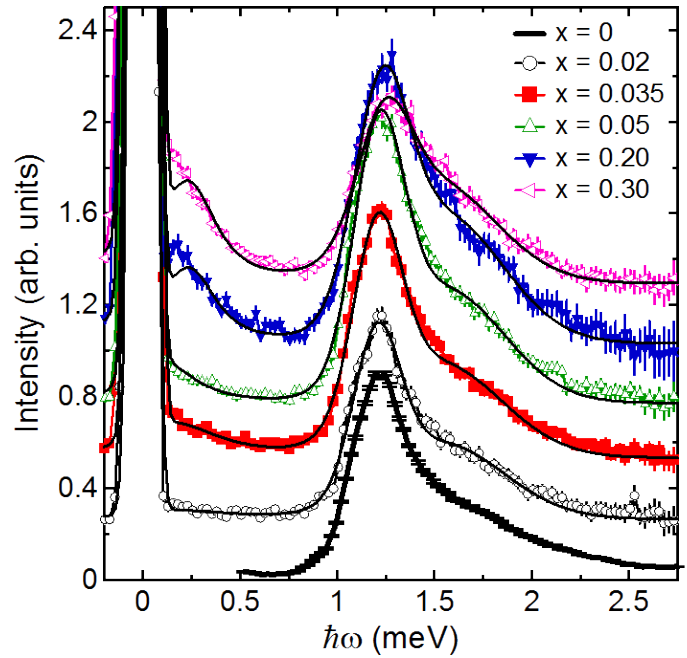

Figure 7. (color online) Inelastic neutron scattering intensity as a function of energy transfer for different concentrations of vanadium in $Ba_3(Mn_{1-x}V_x)_2O_8$. Data have been integrated between $Q$ = 0.6 and 1.7 Å$^{-1}$. Data are offset vertically along the y-axis. Solid lines are a parameterization of the elastic scattering and inelastic scattering as described in the text. The x=0 data are reproduced from Ref. 23.

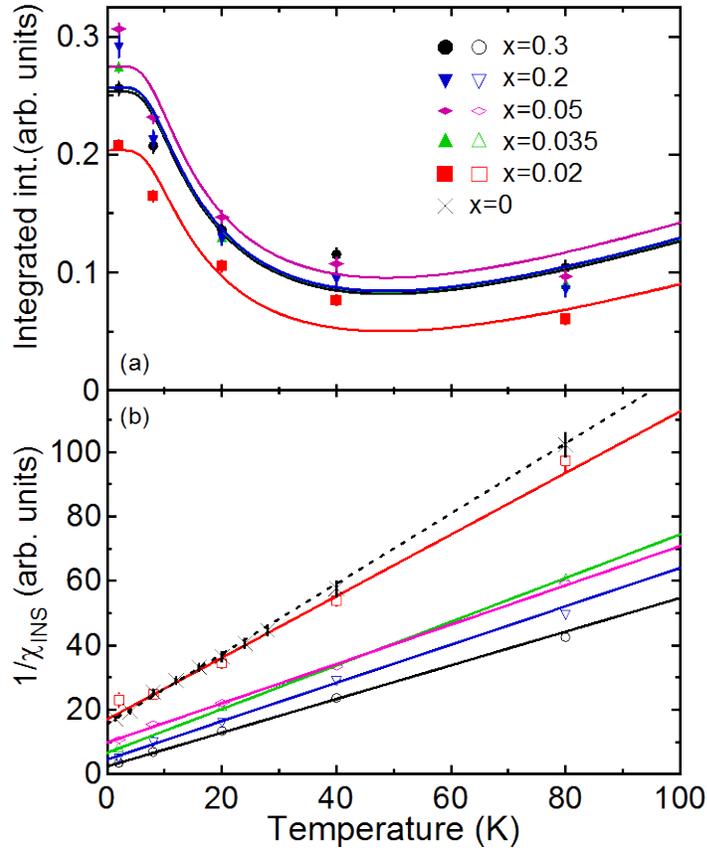

Figure 8. (color online) (a) Integrated inelastic scattering intensity as a function of temperature for different vanadium substitutions, x, in $Ba_3(Mn_{1-x}V_x)_2O_8$. Data have been integrated between 0.5 and 1.5 Å$^{-1}$ wave vector transfer for energy transfer $0.5 < \hbar\omega < 2.9$ meV. Solid lines are a global fit to the dimer model described in the text. (b) The Bose-factor normalized inverse scattering intensity of the quasi-elastic scattering intensity for different x in $Ba_3(Mn_{1-x}V_x)_2O_8$ as a function of temperature. Data have been integrated between 0.5 and 1.5 Å$^{-1}$ wave vector transfer for energy transfers between 0.25 and 0.5 meV. Solid lines are linear fits described in the text. Dashed line is a linear fit to the x=0 data as described in the text.

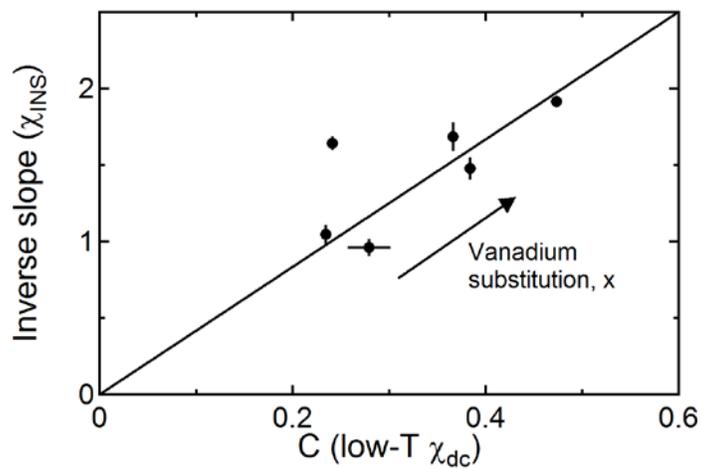

Figure 9. Inverse slope of the fitted lines shown in Figure 6(b) plotted versus the free spin concentration, $C$, shown also in Figure 5(d). Solid line is a linear fit as described in the text. Arrow indicates the direction of increasing vanadium substitution, x in $Ba_3(Mn_{1-x}V_x)_2O_8$, for the data points shown.

[1] B. S. Shastry and B. Sutherland, Phys. Rev. Lett. **47**, 964 (1981).

[2] F. D. M. Haldane, Phys. Rev. B **25**, 4925 (1982).

[3] L. Noodleman, J. Chem. Phys. **74**, 5737 (1981).

[4] T. Giamarchi, Ch. Rüegg, and O. Tchernyshyov, Nature Physics **4**, 198 (2008).

[5] M. B. Stone, C. Broholm, D. H. Reich, O. Tchernyshyov, P. Vorderwisch and N. Harrison, Phys. Rev. Lett. **96**, 257203 (2006).

[6] A. Zheludev, V. O. Garlea, T. Masuda, H. Manaka, L.-P. Regnault, E. Ressouche, B. Grenier, J.-H. Chung, Y. Qiu, K. Habicht, K. Kiefer, and M. Boehm, Phys. Rev. B **76**, 054450 (2007).

[7] B. Leuenberger, H. U. Güdel and J. K. Kjems, J. Mag. Mag. Matt. **53**, 175 (1985).

[8] Ch. Rüegg, B. Normand, M. Matsumoto, A. Furrer, D.F. McMorrow, K. W. Krämer, H.-U. Güdel, S. Gvasaliya, H. Mutka, and M. Boehm, Phys. Rev. Lett. **100**, 205701 (2008).

[9] H. Tanaka, K. Goto, M. Fujisawa, T. Ono, Y. Uwatoko, Physica B: Condensed Matter **329-333B**, 697 (2003).

[10] T. Hong, V. O. Garlea, A. Zheludev, J. A. Fernandez-Baca, H. Manaka, S. Chang, J. B. Leao, and S. J. Poulton, Phys. Rev. B **78**, 224409 (2008).

[11] T. Masuda, A. Fujioka, Y. Uchiyama, I. Tsukada and K. Uchinokura, Phys. Rev. Lett. **80**, 4566 (1998).

[12] L. P. Regnault, J. P. Renard, G. Dhalenne, and A. Revcolevschi, Europhys. Lett. **32**, 579 (1995); Y. Sasago, N. Koide, K. Uchinokura, M. C. Martin, M. Hase, K. Hirota, and G. Shirane, Phys. Rev. B **54**, R6835 (1996); M. C. Martin, M. Hase, K. Hirota, G. Shirane, Y. Sasago, N. Koide, and K. Uchinokura, Phys. Rev. B **56**, 3173 (1997).

[13] T. Masuda, I. Tsukada, K. Uchinokura, Y. J. Wang, V. Kiryukhin, and R. J. Birgeneau, Phys. Rev. B **61**, 4103 (2000).

[14] T. Waki, Y. Itoh, C. Michioka, K. Yoshimura, M. Kato, Phys. Rev. B **73**, 064419 (2006).

[15] S. B. Oseroff, S.-W. Cheong, B. Aktas, M. F. Hundley, Z. Fisk, L. W. Rupp Jr., Phys. Rev. Lett. **74**, 1450 (1995).

[16] M. Azuma, Y. Fujishiro, M. Takano, M. Nohara, and H. Takagi, Phys. Rev. B **55**, R8658 (1997).

[17] Y. Uchiyama, Y. Sasago, I. Tsukada, K. Uchinokura, A. Zheludev, T. Hayashi, N. Miura, and P. Boni, Phys. Rev. Lett. **83**, 632 (1999).

[18] A. Oosawa, T. Ono, and H. Tanaka, Phys. Rev. B **66**, 020405 (2002).

[19] H.-J. Mikeska, A. Ghosh, and A. K. Kolezhuk, Phys. Rev. Lett. **93**, 217204 (2004).

[20] J. Bobroff, N. Laflorencie, L. K. Alexander, A. V. Mahajan, B. Koteswararao, and P. Mendels, Phys. Rev. Lett. **103**, 047201 (2009).

[21] H. Tsujii, B. Andraka, M. Uchida, H. Tanaka, and Y. Takano, Phys. Rev. B **72**, 214434 (2005).

[22] E. C. Samulon, K. A. Al-Hassanieh, Y.-J. Jo, M. C. Shapiro, L. Balicas, C. D. Batista, I. R. Fisher, Phys. Rev. B **81**, 104421 (2010).

[23] M. B. Stone, M. D. Lumsden, Y. Qiu, E. C. Samulon, C. D. Batista, and I. R. Fisher, Phys. Rev. B **77**, 134406 (2008).

[24] M. B. Stone, M. D. Lumsden, S. Chang, E. C. Samulon, C. D. Batista, and I. R. Fisher, Phys. Rev. Lett. **100**, 237201 (2008).

[25] M. B. Stone, M. D. Lumsden, S. Chang, E. C. Samulon, C. D. Batista, and I. R. Fisher, Phys. Rev. Lett. **105**, 169901(E) (2010).

[26] B. Xu, H-T. Wang, and Y. Wang, Phys. Rev. B **77**, 014401 (2008).

[27] A. Oosawa, M. Fujisawa, K. Kakurai, and H. Tanaka, Phys. Rev. B **67**, 184424 (2003).

[28] M. Azuma, M. Takano, and R. S. Eccleston, J. Phys. Soc. Jpn. **67**, 740 (1998).

[29] J. A. Mydosh, *Spin Glasses: An Experimental Introduction*, CRC Press (1993).

[30] E. C. Samulon, M. C. Shapiro and I. R. Fisher, arXiv:1011.6423.

[31] Values in parentheses and error bars shown in figures are the standard deviation.

[32] A. C. Larson and R. B. VonDreele, Los Alamos National Laboratory Report LAUR 86-748 (2004).

[33] M. T. Weller and S. J. Skinner Acta Crystallogr., Sect. C: Cryst. Struct. Commun. **C55**, 154 (1999).

[34] S. J. Mugavero III, M. Bharathy, J. McAlum, H. C. zur Loye, Solid State Sci. **10**, 370 (2008).

[35] G. Liu and J. E. Greedan, J. Solid State Chem. **110**, 274 (1994).

[36] M. Uchida, H. Tanaka, H. Mitamura, F. Ishikawa and T. Goto, Phys. Rev. B **66**, 054429 (2002).

[37] H. Kageyama, K. Yoshimura, R. Stern, N. V. Mushnikov, K. Onizuka, M. Kato, K. Kosuge, C. P. Slichter, T. Goto, and Y. Ueda, Phys. Rev. Lett. **82**, 3168 (1999).

[38] S. Miyahara and K. Ueda, Phys. Rev. Lett. **82** 3701 (1999).

[39] S. Haravifard, S. R. Dunsiger, S. El Shawish, B. D. Gaulin, H. A. Dabkowska, M. T. F. Telling, T. G. Perring, and J. Bonča, Phys. Rev. B **97**, 247206 (2006).

[40] A. A. Aczel, G. J. MacDougall, J. A. Rodriguez, G. M. Luke, P. L. Russo, A. T. Savici, Y. J. Uemura, H. A. Dabkowska, C. R. Wiebe, J. A. Janik, and H. Kageyama, Phys. Rev. B **76**, 214427 (2007).

[41] S. E. Shawish and J. Bonča, Phys. Rev. B **74**, 174420 (2006).

[42] S. Capponi, D. Poiblanc, and F. Mila, Phys. Rev. B **80**, 094407 (2009).

[43] S. Manna, S. Majumder, and S. K. De, J. Phys.: Condens. Matter **21**, 236005 (2009).

[44] S. K. Ma, C. Dasgupta, and C-K. Hu, Phys. Rev. Lett. **43**, 1434 (1979).

[45] R. N. Bhatt and P. A. Lee, Phys. Rev. Lett. **48**, 344 (1982).

[46] D. S. Fisher, Phys. Rev. B **50**, 3799 (1994).

[47] T. R. Kirkpatrick and D. Belitz, Phys. Rev. Lett. **76**, 2571 (1996).

[48] K. S. D. Beach, March meeting of the American Physical Society.

http://meetings.aps.org/link/BAPS.2010.MAR.Y30.6